\documentstyle[11pt]{article}

\begin{document}
\begin{titlepage}
\title{{\bf The vacuum structure in a supersymmetric gauged Nambu-Jona-Lasinio
            model.}}
{\bf
\author{
        P.Bin\'etruy, E.A.Dudas\thanks{Supported in part by the CEC Science
        project no. SC1-CT91-0729}
        , F.Pillon.\\
        Laboratoire de Physique Th\'eorique et Hautes \'Energies\\
        Universit\'e de Paris-Sud, Bat. 211, 91405 Orsay, France
        \thanks{Laboratoire associ\'e au Centre National de la Recherche
        Scientifique}.}
   }
\date{}
\maketitle

\begin{abstract}
\noindent
The dynamical breakdown of the $SU(2) \times U(1)$ symmetry triggered by a
top-antitop condensate is studied in a supersymmetric version of the
gauged Nambu-Jona-Lasinio model. An effective potential approach is used to
investigate the vacuum structure and the equivalence with the minimal
supersymmetric standard model. The role of the soft supersymmetry breaking
terms is analyzed in detail in a version of the model where the electroweak
gauge interactions are turned off.

\end{abstract}
\thispagestyle{empty}
\vfill
LPTHE 93/07 (February 1993).
\end{titlepage}

\newcounter{oldequation}

\def\subequations{\setcounter{oldequation}{\c@equation}%
  \setcounter{equation}{1}%
  \global\let\oldtheequation\theequation%
  \@ifundefined{chapter}%
       {\gdef\theequation{\theoldequation\alph{equation}}
        \global\let\thewholequation\theoldequation
        \gdef\@currentlabel{\theoldequation\alph{equation}}}
       {\gdef\theequation{\thechapter.\theoldequation\alph{equation}}
        \gdef\thewholequation{\thechapter.\theoldequation}
        \gdef\@currentlabel{\thechapter.\theoldequation\alph{equation}}}
  \global\let\@currenteqlabel\thewholequation}

\def\nosubequations{\setcounter{equation}{\c@oldequation}%
  \stepcounter{equation}
  \global\let\theequation\oldtheequation
  \global\let\@currentlabel\theequation}

\def\newsubequation{\stepcounter{oldequation}
    \setcounter{equation}{1}}

\def\@currenteqlabel{}

\def\eqlabel#1{\@bsphack\if@filesw {\let\thepage\relax
   \xdef\@gtempa{\write\@auxout{\string
      \newlabel{#1}{{\@currenteqlabel}{\thepage}}}}}\@gtempa
   \if@nobreak \ifvmode\nobreak\fi\fi\fi\@esphack}

\section{Introduction.}

      The idea of a dynamical symmetry breaking mechanism at work in the
standard model is appealing because it requires no fundamental scalar field to
exist. Technicolor \cite{first}, a priori the most attractive approach, assumes
that a new QCD-type strong interaction produces bound states at a scale of the
order of the electroweak breaking scale; the interaction of the $SU(2)\times
U(1)$ gauge fields with these composite fields reproduces the usual mass
spectrum and mixing angle in the gauge sector.
Unfortunately, problems arise when one tries to give masses to the usual
fermions while avoiding FCNC processes. This requires substantial modifications
leading to not very economic models like Walking Technicolor \cite{ab}
which, moreover, hide a yet unknown dynamics.
Even taking into account recent progress \cite{ac} it is not yet clear that
one has an
explicit example that solves unambigously all the problems.

    One can search for other possibilities such as dynamical symmetry breaking
mechanisms of the Nambu-Jona-Lasinio type which are more tractable at a
computational level but are associated with non renormalisable couplings: one
is
forced to introduce a physical cut-off $\Lambda$ which cannot be eliminated
\cite{ad}.
Applied to the standard model this yields a composite Higgs particle viewed as
a
top-antitop bound state which reproduces the usual low energy phenomenology
\cite{ae}.
The infrared equivalence with the usual standard model has been proved in a
$\frac{1}{N_c}$ expansion \cite{af} and it has been noted that this theory
lacks predictive power due to higher-dimension operators which come into play
when one approaches the compositeness scale $\Lambda_c$ where the Higgs
bound state dissociates  \cite{afa} (see however Ref.\cite{ag}).

     The central problem of giving fermions a mass while avoiding flavor
changing neutral current processes,
may be solved \cite{ah}. Moreover we do not have the usual
tower of pseudo Goldstone
bosons of technicolor models. Unfortunately, such an approach does not avoid
the fine-tuning problem: to obtain a light fermion mass (compared to the
cut-off scale), one must fix the four-fermion coupling with an extraordinary
precision. As usual , one possibility to evade this problem is to make the
model
supersymmetric \cite{ai}: quadratic divergences cancel and the fermions
can be made {\it naturally} light. Of course we lose the original motivation
which was to avoid elementary scalar fields. On the other hand, the
scalar fields which
are thus introduced are much heavier than the usual Higgs particles so that we
can keep part of the motivation in saying that there are no elementary
scalar degree of freedom at the presently accessible energies (i.e. electroweak
unification).

    One would like to be able to explicitly check the infrared equivalence
with the Minimal Supersymmetric Standard Model (MSSM) which was assumed in
ref. \cite{ai}, where the gap equation was
written in superfield language in a leading $\frac{1}{N_c}$ expansion.

In this respect, a first point of relevance is the role played by the soft
supersymmetry breaking terms. It has been noticed for some time \cite{aj}
that their presence is necessary in order to have chiral symmetry breaking.
Otherwise, supersymmetry would have to be spontaneously broken and a chiral
symmetry breaking ground state would be at best degenerate with the trivial
vacuum. From the MSSM point of view, one soft supersymmetry breaking term
plays a key role: it is the one that yields a mixing between the two Higgs
doublets. It is usually written as $B \mu H_1 \cdot H_2 + h.c.$, where $\mu$
is the supersymmetric mass term present in the superpotential. If this term is
not present, the theory is plagued with an unwanted axion and zero v.e.v. for
$H_1$ or $H_2$. The supersymmetric parameter $\mu$
appears naturally in the linearized
version of the supersymmetric Nambu-Jona-Lasinio model. On the other hand, $B$,
as any other supersymmetry breaking parameter, is put by hand: one invokes for
that an underlying supergravity theory, whose precise content is not specified.
We will study below in detail the role played by this parameter $B$ in the
chiral symmetry breaking.

To check the equivalence, it is also of importance to include the effect
of {\it electroweak} gauge interactions through the auxiliary fields in the
gauge sector (D-terms). Indeed, in the MSSM, the electroweak gauge
couplings are necessary to get the symmetry breaking pattern of $SU(2)\times
U(1)$: they give the only quartic interactions in the scalar potential.
Hence, neglecting them means either vanishing Higgs vacuum expectation values
or unbounded potential.
These couplings must therefore be taken into account when checking the
equivalence. This is most easily done in the component formalism where the
relevant D-terms can readily be isolated (by contrast with the superfield
approach).

      Our approach consists in writing an effective potential for the composite
Higgs fields in the leading $\frac{1}{N}$ expansion, in the spirit of refs.
\cite{af} and \cite{ak}.
We then investigate the vacuum structure and the possible second order phase
transitions in the space of the parameters appearing in the lagrangian. In
order to do so, a gauged
supersymmetric version of a Nambu-Jona-Lasinio model is written in a linearized
form introducing two auxiliary fields $H_1$ and $H_2$. Through radiative
corrections they acquire a dynamics, becoming propagating degrees of freedom.
Their quantum numbers match with those of the Higgs fields in the MSSM. The
effective potential for the scalar components of $H_1$ and $H_2$, in leading
$\frac{1}{N}$, is written and the saddle point equations analyzed to find
the real vacuum.

Section 2 gives a short presentation of the $\frac{1}{N}$ expansion as needed
in
what follows. In section 3, we analyze the role of the soft supersymmetry
breaking parameters, more particularly of the $B$ term which plays an important
role in the MSSM. We study the dependance on $B$ of the critical parameters
of the second order phase transition associated with the breaking of chiral
symmetry in a non-gauged version of the model. This also allows us to test
the efficiency of the method that we will use later on the gauged model.
Section 4 derives the $\beta$ function and discusses the symmetries
together with the scalar spectrum for this non-gauged version.
Section 5 deals with the vacuum structure
of the gauged model; a detailed discussion of the equivalence with the MSSM is
presented. Finally, Section 6 concludes. Some computational details are
also given in an appendix.

\section{The 1/N expansion.}

      Non-perturbative summation methods for Feynman graphs like the semi
classical development and the $\frac{1}{N}$ expansion give important
information about the pattern of symmetry breaking, non trivial fixed points
in the renormalisation group evolution, bound states formation, etc..
The $\hbar$
semiclassical development corresponds to an expansion in the number of loops
and
the first one-loop order is easily written in a compact form. This is to be
contrasted with the $\frac{1}{N}$ expansion, especially in the gauge field
sector of an
non-abelian theory where a resummation to a given order cannot be performed
analytically. We may easily understand the qualitative difference by trying to
give a criterion for computing the leading order in the two cases:

  (a) In the semiclassical $\hbar$ development, one uses the partition
function:

$$        Z = \int {\cal D}\Phi e^{{i \over \hbar} S[\Phi]}       $$
where  $\Phi$ is a set of quantum fields and S[$\Phi$] is the action of the
general model considered. The order of a given Feynman graph is given by:

$$ \hbar^{I-V} = \hbar ^ {L-1} $$
where L is the number of loop(s) in the Feynman graph, this formula being
valid irrespective of the dynamics of the model. In the leading order,
the vacuum structure  can be studied using the Coleman-Weinberg effective
potential \cite{al}.

  (b) The $\frac{1}{N}$ expansion is model dependent.
Take, for example, a Gross-Neveu
U(N) theory with N fermions \cite{ak} and study the dynamical breakdown of the
chiral symmetry. Introducing an auxiliary scalar field, the Lagrangian is
given by ($g_0$ is independent of $N$):

\begin{eqnarray}
{\cal L} & = & {{\bar{\Psi}^a}} i {{\gamma}^{\mu}} {\partial_\mu} {\Psi^a}
+ {g_0 \over 2N} {({{{\bar\Psi}^a}}{\Psi^a}) ^2} - {N\over {2g_0}}
{ (\sigma - {g_0 \over N} ( { {\bar\Psi}^a}  {\Psi^a} )) ^2 } \cr
& = & {{\bar\Psi}^a} i {{\gamma}^{\mu}} {\partial_\mu} {\Psi^a}
- {N \over 2g_0} {\sigma^2} + {\sigma} { {\bar\Psi}^a} {\Psi^a} \cr
& = & N ({\bar \Psi}^{\prime a} i {\gamma^\mu} {\partial_\mu}
{\Psi^{\prime a}} - {1 \over 2g_0}{\sigma^2} + {\sigma}
{\bar \Psi}^{\prime ^a} {\Psi^{\prime a}} )
\end{eqnarray}
where in the last line we have performed the rescaling $\Psi \longrightarrow
\sqrt{N} \Psi'$, which modifies the partition function with an irrelevant
constant. The main difference with respect to the loop ($\hbar$) expansion
is that every closed fermion loop gives an additional $N$ factor. So, for
an arbitrary graph, the associated factor is
$$ N^{-I+V+{L_f}} = N^{ 1- (L-L_f)} $$
where $L_f$ is the number of fermion loops $(L_f < L)$.
The leading order corresponds to $L-L_f = 0$. More precisely, in the model that
we consider, it is $L = L_f = 1$. Indeed, in any one-particle irreducible
diagram, the introduction of scalar internal lines make the diagram
subleading. Thus, the leading diagrams correspond to a single fermion loop and
external scalar fields $\sigma$ only. The effective potential for the $\sigma$
field is exactly given in the leading order by the Coleman-Weinberg expression,
or equivalently in the Jackiw functional form \cite{al}:

$$e^{i S(\sigma_c)} = \int {{\cal D}({{\overline \Psi}^a},{\Psi^a},{\sigma_q})}
\exp{S_{quadratic}(\sigma_c + \sigma_q ,\overline {\Psi}^a,\Psi^a)}.$$
Several points actually depend on the model considered:

   -The factor associated with fermion loops (N) depends on the group
representation.

   -The leading order condition $L-L_f$ = minimum, has solutions which
depend on the interaction lagrangian (e.g. gauge fields).

   -More generally, it is not clear whether every model has a non
trivial $\frac{1}{N}$ development.

   -If we want to apply the effective potential computation to the ground state
determination, we may find an N dependent vacuum expectation value which will
change the leading order contribution. This corresponds, in terms of Feynman
graphs to a multiplication of every external leg with a N dependent factor.

  Being interested in the dynamical breaking of the chiral symmetry in
supersymmetric models, we study in the leading order in N the linearised
version of the supersymmetrized Gross-Neveu lagrangian
\cite{ai}. This lagrangian is
written in a superfield language after performing the superfield rescaling
\begin{eqnarray}
{\Phi_i} \longrightarrow \sqrt{N} \Phi_i  \: , \: H_{1,2}  \longrightarrow
\sqrt{N}  H_{1,2}
\end{eqnarray}
as
\begin{eqnarray}
{\cal L}  & = &
N \left\{ \int {d^4}\Theta \: [{\Phi_i^+}{\Phi_i}(1-{\Sigma^2}{\Theta^2}
{{\bar \Theta}^2})] \: + \: {H_1^+} {H_1} (1 - {\Delta^2} {\Theta^2}{{\bar
\Theta}^2}) \right. \cr
& & +  \int {d^2} \Theta \: {H_2} (m {H_1} - g_0 {\Phi_i}{\Phi_i})
+ \int {d^2} {\bar \Theta} \: {H_2^+} (m{H_1^+} - g_0 {\Phi_i^+}{\Phi_i^+}) \cr
& & + \left.\: {\int {d^2} \Theta} \: {H_2} {H_1} Bm {\Theta^2}
+ {\int {d^2} {\bar \Theta}} {H_2^+} {H_1^+} Bm{{\bar \Theta} ^2} \right\}
\end{eqnarray}
where i runs from 1 to N. We have included soft supersymmetry breaking terms
$(\Sigma \: , \: \Delta \:$ and $\: B)$.
The same analysis as above yields for a  given supergraph a factor:
$$ N^{-I+V+{L_{\Phi_i}}} = N^{1-(L-L_{\Phi_i})}.$$
The leading N contribution is given by $L-L_{\Phi_i}$ minimum which again
means $L = L_{\Phi_i} = 1$, that is no $\Phi_i$ on the external legs.

Consequently, writing the effective potential {\cal V}($H_1$,$H_2$) in the
leading N expansion is equivalent to using the Coleman-Weinberg one-loop
formula,
taken care of the fact that the different v.e.v. of the different fields do
not change this result. As we will explicitly verify, it does not happen
to be the case here but it does occur in non-minimal supersymmetric
generalisations of the Gross-Neveu model.

If we gauge the lagrangians (1) or (3), the analysis is different,
due to the non-abelian structure of the symmetry group. As is well-known,
the leading contribution is then given by all planar diagrams \cite{am}.
However, we are not so much interested here  in gauging the degrees of freedom
that correspond to the strong interactions as in including the effect of
electroweak $SU(2) \otimes U(1)$ gauge interactions. Therefore, when we deal
with this question in section 5, we will introduce $SU(2) \otimes U(1)$
propagating gauge degrees of freedom whereas we will still
consider QCD gauge degrees of freedom as some sort of flavor number (taking
N values).

For the time being, we will consider the simpler case where no gauge degrees
of freedom are propagating and we will get a closer look at the role played by
soft supersymmetry breaking terms.

\vskip 1cm

\section{Dynamical chiral symmetry breaking (DCSB) in the minimal
supersymmetric Gross-Neveu model in the 1/N expansion.}

The supersymmetric Nambu-Jona-Lasinio model is readily written in its
simplest form as:
\begin{eqnarray}
{\cal L} = \int d^4\Theta [ \Phi_i^+ \Phi_i + G \ \Phi_i^+ \Phi_i
\Phi_j^+ \Phi_j].
\end{eqnarray}
It has been immediately realized \cite{aj} that, in order to write a linearized
version of this model, one needs two chiral superfields $H_1$ and $H_2$ whose
couplings are described by eq.(3) with $G=g^2/m^2$ (eq.(3) includes also,
as emphasized above, soft supersymmetry breaking terms).
This doubling of the auxiliary fields
\footnote{It is important to note that, despite a kinetic term for $H_1$
in (3), both superfields $H_1$ and $H_2$ are in fact composite fields
which can be eliminated through their equations of motion.}
is a welcome feature
since it also reproduces some of the couplings of the MSSM \cite{ai}.
Indeed, one naturally finds in (3) a supersymmetric mass term $m H_1 H_2$
mixing the two Higgs. This is reminiscent of the $\mu$ parameter of the MSSM.
In the MSSM, the presence of the $\mu$ term is a problem (the so-called $\mu$
problem\cite{an}) because $\mu$ is naturally either zero or of the order of the
underlying scale, whose role is played here by our cut-off $\Lambda$. We will
return to this problem shortly.

We wish to study in detail in this section the role of the soft supersymmetry
breaking terms introduced in (3). As noticed earlier \cite{aj,ai} and
emphasized above, such terms -- in fact the $\Sigma$ mass term --
are necessary in order to get a chiral symmetry breaking ground state.
We will rederive this result but will be more interested in the role played
by the $B$ parameter which is so important in the MSSM. We will in particular
study how the second-order phase transition depends on this parameter.

The dynamics resulting from lagrangian (3)
can be studied diagrammatically in relation with DCSB \cite{ai}.
Alternatively, we will compute by functional methods the effective potential
for $H_1$ and $H_2$ and minimize it. A nonzero value for $<H_1>$ or $<H_2>$
breaks a U(1) chiral symmetry which will be precisely defined in the
next section.

In components, the lagrangian (3) reads\footnote{We have undone the rescaling
(2) and redefined the coupling g in order to include the N dependence: $g=
g_0 / \sqrt{N}$.}:

\begin{eqnarray}
{\cal L} & = &  {F^*_1}{F_1} + { z^*_1}(\Box - {\Delta^2}){z_1}
- i{{\bar \Psi}_1}{{\bar \sigma}^m}{\partial _m}{\Psi _1}
+ {F^*_i}{F_i} + {z^*_i}(\Box - {\sigma^2}){z_i} \cr
& & - i{\bar{\Psi}_i}{{\bar \sigma}^m}{\partial_m}{\Psi_i}
+ m({z_1}{F_2}+{z_2}{F_1} - {\Psi_1}{\Psi_2} - {{\bar \Psi}_1}{{\bar \Psi}_2})
\cr
& & - g({F_2}{z_i}{z_i} + 2{F_i}{z_2}{z_i} - {z_2}{\Psi_i}{\Psi_i}
- 2{z_i}{\Psi_i}{\Psi_2} + h.c.) \cr
& & - Bm({z_1}{z_2} + {z^*_1}{z^*_2}).
\end{eqnarray}

Integrating over $F_i$ (we will verify that $<F_i> = 0$) and using the
formula for the effective potential in the leading one-loop order:

\begin{eqnarray}
V_{eff} = V_{tree} + \frac{1}{2} {\cal{ST}}r \int \frac{{d^d}p}{(2 \pi)^d}
\ln ( p^2 + M^2 )
\end{eqnarray}
where ${\cal{ST}}r F( M^2 ) = Tr ( M_s^2 ) - 2 Tr ( M_f^2 )$, we readily obtain

\begin{eqnarray}
V_{eff} & = & V_{tree} + \frac{N}{2} \int \frac{{d^d}p}{(2 \pi)^d} [\ln ( p^2
+ \Sigma^2 + 4g^2 |{z_2}|^2 + 2g \sqrt{F_2 F_2^*}) \cr
& & + \ln ( p^2 + \Sigma^2 + 4g^2 |{z_2}|^2 - 2g \sqrt{F_2 F_2^*})
- 2 \ln (p^2 + 4g^2 |z_2|^2)]
\end{eqnarray}

As noticed earlier, there are no one-loop contribution to the $z_i$-dependent
part of the potential. Minimizing with respect to $z_i^* , F_1^* , z_1^* ,
F_2^*$ and $z_2^*$ yields the following equations:

\begin{eqnarray}
z_i = 0
\end{eqnarray}
\begin{eqnarray}
F_1 + m z_2^* & = & 0
\end{eqnarray}
\begin{eqnarray}
\Delta^2 z_1 - m F_2^* - Bmz_2^* & = & 0
\end{eqnarray}
\begin{eqnarray}
m z_1^* + 2 g^2 F_2 N \int \frac{{d^d}p}{(2 \pi)^d} \frac{1}
{( p^2 + \Sigma^2 + 4g^2 |{z_2}|^2)^2 - 4g^2 {F_2 F_2^*}} = 0
\end{eqnarray}
\begin{eqnarray}
4g^2 z_2 N \int \frac{{d^d}p}{(2 \pi)^d} \left[ \frac{p^2
+ \Sigma^2 + 4g^2 |{z_2}|^2}{( p^2 + \Sigma^2
+ 4g^2 |{z_2}|^2)^2 - 4g^2 {F_2 F_2^*}}
- \frac{1}{p^2 + 4g^2 |z_2|^2} \right] \nonumber
\end{eqnarray}
\begin{eqnarray}
- m F^*_1 -Bmz_1^*  = 0
\end{eqnarray}

Since $F_i^* = 2g z_2 z_i$, we obtain the announced result $F_i = 0$. Combining
the four non trivial equations we find the system:

\begin{eqnarray}
B \frac{m^2}{\Delta^2} z_2 =
- F_2 \left[ \frac{m^2}{\Delta^2} + 2 g^2 N \int \frac{{d^d}p}{(2 \pi)^d}
\frac{1}{(( p^2 + \Sigma^2 + 4g^2 |{z_2}|^2)^2 - 4g^2 {F_2 F_2^*})} \right]
\end{eqnarray}

\begin{eqnarray}
B\frac{m^2}{\Delta^2} F_2 & = & z_2 \left\{  4g^2 N \int\frac{{d^d}p}{(2\pi)^d}
\left[ \frac{p^2 + \Sigma^2 + 4g^2 |{z_2}|^2}{(p^2+\Sigma^2 +4g^2 |{z_2}|^2)^2
- 4g^2 {F_2 F_2^*}} \right. \right. \cr
& & \left. \left. -\frac{1}{p^2 + 4g^2 |z_2|^2}  \right]
+ m^2 (1-\frac{B^2}{\Delta^2}) \right\}
\end{eqnarray}

Of course, because the original theory is non-renormalisable, all integrals
have to be cut off at a scale $\Lambda$.
For future use  we give the explicit form in four dimensions:

\begin{eqnarray}
& & B \frac{m^2}{\Delta^2} z_2 = F_2 \: \left[ \: - \frac{m^2}{\Delta^2}
+ \frac{Ng^2}{16\pi^2} [ \ln{ \frac{(\Sigma^2 + 4g^2 |{z_2}|^2)^2
- 4g^2 {F_2 F_2^*}}{\Lambda^4}} \right. \cr
& & + \left.  {\frac{1}{2g \sqrt{F_2^* F_2} }} (\Sigma^2 + 4g^2 |{z_2}|^2)
\ln {\frac{\Sigma^2 + 4g^2 |{z_2}|^2 + 2g \sqrt{F_2 F_2^*}}{\Sigma^2
+ 4g^2 |{z_2}|^2 - 2g \sqrt{F_2 F_2^*}}} \: \right] \cr
& & + {\cal O}(\frac{1}{\Lambda^2})
\end{eqnarray}
\begin{eqnarray}
& & \frac{m^2}{\Delta^2}F_2 = z_2 \left\{
\frac{Ng^2}{8\pi^2} \left[ (\Sigma^2 + 4g^2 |{z_2}|^2)\ln{\frac{(\Sigma^2
+ 4g^2 |{z_2}|^2)^2 - 4g^2 {F_2 F_2^*}}{\Lambda^4}} \right. \right. \cr
& & \left. \left. -8g^2 |z_2|^2 \ln {\frac{8g^2 |z_2|^2 }{\Lambda^2}}
+ 2g \sqrt{F_2 F_2^*} \ln{\frac{\Sigma^2 + 4g^2 |{z_2}|^2 + 2g \sqrt
{F_2 F_2^*}}{\Sigma^2 + 4g^2 |{z_2}|^2 - 2g \sqrt{F_2 F_2^*}}} \right]
\right. \cr
& & \left. + m^2 (1 - \frac{B^2}{\Delta^2}) \right\}
+ {\cal O}(\frac{1}{\Lambda^2})
\end{eqnarray}

The equations can easily be recast in a form which was already obtained by
Carena {\it et al.}\cite{ai} (see appendix, section iii)).

The case B=0 has been studied in detail by Clark, Love and Bardeen (CLB) in
ref. \cite{ai}. In agreement with their analysis, we immediately obtain from
equations (11) and (13) that $F_2 = z_1 = 0$; equation (14) then reads:

\begin{eqnarray}
z_2 \left[ m^2 - 4g^2N\Sigma^2 \int^{\Lambda}
{\frac{{d^d}p}{(2 \pi)^d} \frac{1}{(p^2 + \Sigma^2
+ 4g^2 |{z_2}|^2)( p^2 + 4g^2 |{z_2}|^2)}} \right] =0
\end{eqnarray}
which has a non trivial solution $<z_2> \neq 0$ for $G \equiv g^2/m^2 >G_c$,
with $G_c$ defined by the equation:

\begin{eqnarray}
G_c^{-1} = 4N \Sigma^2 \int^{\Lambda} {\frac{{d^d}p}{(2 \pi)^d} \frac{1}
{p^2(p^2 + \Sigma^2)}}\ .
\end{eqnarray}
This gives $G_c^{-1} = {N\Sigma^2 \over 4 \pi^2} \ln {\Lambda^2 \over
\Sigma^2}$
in $d=4$ spacetime dimensions.

We can readily verify that this corresponds to the true ground state of
the theory for $G<G_c$ because, using the saddle point equations, one
can derive:

\begin{eqnarray}
V(<z_2>) - V(0) = \frac{N\Sigma^4}{32\pi^2} f(\frac{4g^2
|{<z_2>}|^2}{\Sigma^2})
\end{eqnarray}
where, for $d=4$,
$f(x) \equiv (1-x^2)\ln{(1+x)} + x^2 (\ln{x} -1) < 0$.

Before leaving this simple case, let us note the form of the nontrivial
solution of (17) in the case of $d=4$ spacetime dimensions and a large cut-off
$\Lambda$:
\begin{eqnarray}
G^{-1}={m^2 \over g^2}={N \over 4\pi^2}
\left[\Sigma^2\ln{\frac{\Lambda^2}{\Sigma^2 + 4 g^2 |z_2|^2}}
- 4 g^2 |z_2|^2 \ln{\frac{\Sigma^2 + 4 g^2 |z_2|^2}{4 g^2|z_2|^2}} \right]
\end{eqnarray}
Due to supersymmetry cancellations, the quadratic divergence familiar in
the Nambu-Jona-Lasinio model has cancelled and we are left with logarithmic
divergences only \cite{aj,ai}. This avoids an untolerable fine-tuning but has
the following drawback. The original coupling $G$ behaves typically as
${1 \over \Sigma^2 \ln(\Lambda^2/\Sigma^2)}$ which is surprisingly large
for a theory whose basic scale is $\Lambda$ (one would expect
$G=O({1\over\Lambda^2})$). This is in fact nothing but
the old $\mu$-problem of the MSSM which shows up here in a different language.
Indeed, $G^{-1}$ being small is just a reflexion of the fact that the
parameter $m$ is unnaturally small compared to $\Lambda$ in this theory.

The analysis in the case where the soft symmetry breaking term $B$ is non-zero
is more involved. We first sketch the principles of the method that we use.
Suppose that, for a critical value of the coupling $G=G_c$, we have a second
order (i.e. continuous) phase transition, similar to the one found when $B=0$,
from the trivial vacuum $z_2=F_2=0$ to a non trivial one. In the broken phase,
close to the critical point, i.e. for $G=G_c(1+\epsilon) , 0<\epsilon \ll 1$,
both $z_2$ and $F_2$ must be small:
\begin{eqnarray}
\frac{4g^2 |{z_2}|^2}{\Sigma^2} \equiv \epsilon x_1 \:,\:
\frac{ g^2 |F_2|^2}{\Sigma^4} \equiv \epsilon x_2
\end{eqnarray}
where $x_1$ and $x_2$ are at most of order 1. Introducing these expressions in
the saddle point equations, we can solve them order by order in $\epsilon$.

To the order $\epsilon^0$, we find an equation for $G_c$ which is precisely the
condition that the transition from the trivial vacuum to the nontrivial one is
continuous (second order). We can solve it for $G_c$.

To the order $\epsilon$, we find the behaviour of the order parameter $z_2$ or
$F_2$ in terms of the control parameter $\epsilon = \frac{G-G_c}{G_c}$, from
which we can extract the critical exponents.

The linearized equations (15) and (16) read, to the order $\epsilon$,
\begin{eqnarray}
F_2 \left\{ - \frac{1}{\Delta^2} + \frac{NG_c}{16\pi^2} (1 + \epsilon)
[2 - 4\ln{\frac{\Lambda}{\Sigma}} + \frac{16 g^2 |z_2|^2}{\Sigma^2}
- \frac{4g^2 |F_2|^2}{\Sigma^4}] \right\} = \frac{B}{\Delta^2} z_2,
\end{eqnarray}
\begin{eqnarray}
z_2 \left\{ - \frac{NG_c \Sigma^2}{2\pi^2}
(1 + \epsilon)
[( 1 + \frac{4 g^2 |z_2|^2}{\Sigma^2})    (\frac{ -2 g^2 |z_2|^2}{\Sigma^2}
+ \frac{g^2 |F_2|^2}{\Sigma^4} + \ln{\frac{\Lambda}{\Sigma} }) \right. \cr
- \left.\frac{2g^2 |z_2|^2}{\Sigma^2} \ln {\frac{\Lambda^2}{4g^2 |z_2|^2 }} -
\frac{2g^2 |F_2|^2}{\Sigma^4}] + 1 - \frac{B^2}{\Delta^2} \right\} =
\frac{B}{\Delta^2} F_2.
\end{eqnarray}
Of course, a consistency check on the validity of our approximation will have
to be performed a posteriori.

To the order $\epsilon^0$, we find:
\begin{eqnarray}
\frac{B^2}{\Delta^4} =
[-\frac{1}{\Delta^2} + \frac{NG_c}{16 \pi^2} (2 - 4 \ln
{\frac{\Lambda}{\Sigma}})] [1 - \frac{B^2}{\Delta^2} -
\frac{NG_c\Sigma^2}{2\pi^2} \ln{\frac {\Lambda}{\Sigma}}]
\end{eqnarray}
which reduces to (18) in the limit $B = 0$. Solving (24) and retaining only the
leading terms\footnote{Having in mind phenomenological applications, we can
take for example $\Lambda \simeq 10^{16}$ GeV and $\Sigma \simeq 1$ TeV.}
in $\ln{\frac{\Lambda}{\Sigma}}$, we obtain a unique positive solution given
by:

\begin{eqnarray}
\frac{NG_c}{8\pi^2} = \frac{(\Delta^2 - 2\Sigma^2 - B^2)
+ \sqrt{(\Delta^2 - 2\Sigma^2 - B^2)^2 + 8\Sigma^2 \Delta^2}}{8\Sigma^2
\Delta^2 \ln{\frac{\Lambda}{\Sigma}}}
\end{eqnarray}

We note that $G_c$ decreases with B. Hence $G_c(B \neq 0) < G_c(B = 0)$.
The maximum value for B (beyond which the effective potential becomes
unbounded) is $B^2 = \Delta^2$, in which case we obtain:
\begin{eqnarray}
(G_c)_{min} = \frac{1}{4\Delta^2 \ln{\frac{\Lambda}{\Sigma}}} [ \sqrt{1 +
\frac{2\Delta^2}{\Sigma^2}} - 1]
\end{eqnarray}
The behavior of $G_c$ as a function of B is schematically represented on
Fig.1.
\begin{figure}[p]
\vspace{10cm}
\caption{The critical coupling $G_c$ as a function of B.}
\end{figure}

To study the behaviour of the condensates near the critical point, we take into
account the terms of order $\epsilon$ (and for that matter $\epsilon \ln{
\epsilon}$) in equation (22) and (23).

Defining $\xi_1 \equiv 4g^2 |z_2|^2 / \Sigma^2$ and $\xi_2 \equiv
g^2 |F_2|^2 / \Sigma^4$ and combining (22) and (23), one obtains an
equation of the form:

\begin{eqnarray}
\xi_2 = -x\epsilon + y\xi_1 - z\xi_1 \ln{\xi_1}
\end{eqnarray}
where the quantities $x$, $y$ and $z$ are strictly positive. Using (23) and
the fact that $\epsilon, \: \xi_1 ,\: \xi_2 \: \ll \: 1$, we get in the
leading order:

\begin{eqnarray}
\xi_1 [ (1 - \frac{B^2}{\Delta^2} - \frac{NG_c \Sigma^2}{2\pi^2}
\ln{ \frac{\Lambda}{\Sigma}})^2
- \frac{B^2 \Sigma^2}{\Delta^4}y]
+ \frac{B^2 \Sigma^2}{\Delta^4}z \xi_1 \ln{\xi_1}
= - \frac{B^2 \Sigma^2}{\Delta^4}x \epsilon
\end{eqnarray}
For $\xi_1 \longrightarrow 0$, $\xi_1 \ll \xi_1 |\ln{\xi_1}|$. The first term
of the left-hand side is negligible and one can check that (28) has a
nontrivial solution $\xi_1 \neq 0$ for $\epsilon >0$ only ($\ln{\xi_1}$ is
negative). One finds explicitly:
\begin{eqnarray}
\frac{2B^2 \Sigma^2}{\Delta^4} \xi_1 \ln{\xi_1} &
= & [ (2 \ln{\Lambda}{\Sigma} -1)
(1 - \frac{B^2}{\Delta^2} - \frac{NG_c \Sigma^2}{2\pi^2} \ln{
\frac{\Lambda}{\Sigma}})^2 \cr
& & + \frac{4B^2 \Sigma^2}{\Delta^4} \ln{\frac{\Lambda}{\Sigma}}] \epsilon
\end{eqnarray}
The non zero v.e.v. for $z_1$ is obtained from (29) and the value of $\xi_2$,
using the equation (10). One can now check a posteriori that all the terms
discarded to obtain the result (31) were of higher order.

As always in the mean field approximation for a second order phase transition,
the critical coefficient $\beta$ is $\frac{1}{2}$ \cite{af}. A way to show
this is to perform the computation in $d = 4 + \eta$ dimensions and to take the
limit $\eta \rightarrow 0$ in the expression for $\beta$.
The property that identifies a phase transition
in this model at $G=G_c$, is that
for $G < G_c (\epsilon < 0)$ , $\xi_1 = \xi_2 = 0$ i.e. no condensate can
be formed. It is second order because the condensates appear in a
continuous way from a zero value at $G_c$, as the Nambu-Jona-Lasinio model was
precisely designed for.

\vskip 1cm

\section{The $\beta$ function, symmetries and the mass spectrum.}

\vskip .4cm
Our interest in this section is twofold. First, derive the $\beta$ function
for the model described in the last section. Second, study its symmetries and
the corresponding scalar mass spectrum. This will prove to be useful
in the next section when we undertake to gauge the electroweak degrees of
freedom. It will allow us to compare the vacuum structure of the gauged and
non-gauged model. Since this structure is richer  when $B=0$,
we restrict here our analysis of the symmetries to this case.

\vskip .3cm

\subsection{The $\beta$ function.}

To compute the $\beta$ function, we need only to consider the renormalized
parameters $g$, $m$ and $G = g^2/ m^2$ as they appear from
the one-loop effective potential . The ultraviolet divergences
are
summarized in four dimensions by the following counterterm:
\begin{eqnarray}
{\cal L}_{ct} = -\frac{1}{64 \pi^2} ({\cal{ST}}r M^4) \ln{\Lambda^2}
\cr\cr
{\cal {ST}}r M^4 = N ( 2 \Sigma^4 +16 g^2 |z_2|^2 \Sigma^2 + 8 g^2 |F_2|^2 ).
\end{eqnarray}

This amounts to redefining the tree level mass parameter as follows:
\begin{eqnarray}
m^2 \longrightarrow m^2 ( 1 + \frac{N G}{4\pi^2} \Sigma^2 \ln{\Lambda^2})
\end{eqnarray}
whereas there is no counterterm associated with $g$ to the leading order in
$N$ (see section 3).

Also, using the fact that the term $|F_2|^2$ is a kinetic term for the $H_2$
auxiliary field, one infers from (30) the wave function renormalisation
constant for $z_2$
\begin{eqnarray}
{\cal Z} = 1 -  \frac{N g^2}{8 \pi^2} \ln{\Lambda^2}.
\end{eqnarray}
However the redefinition $z_2^\prime = {\cal Z}^{\frac{1}{2}} z_2$
modifies both couplings $g$ and $m$ with the same factor
${\cal Z}^{-\frac{1}{2}}$, which cancels in
$G = g^2 / m^2$. Therefore only (31) will contribute to the renormalisation
of $g$ which  gives:

\begin{eqnarray}
\beta(G) = {\partial G \over \partial \ln \Lambda}
               = - \frac{N\Sigma^2}{2 \pi^2} G^2.
\end{eqnarray}

So the theory is asymptotically free. More generally, the computation of
the $\beta$ function can be performed in $d = 4 + \eta$ dimensions
using the fact that the B term does not affect the ultraviolet properties
in the leading N order. Then, setting $B = 0$, we derive the gap
equation obtained from (17) and (18) ($\hat{g}^2 \equiv g^2
\Lambda^{-\eta}$)
\begin{eqnarray}
G_c^{-1} - G^{-1} = 4N\Sigma^2 \int^{\Lambda} {d^d p \over (2\pi)^d}
{4 \hat{g}^2 |z_2|^2 ( 2p^2 + \Sigma^2 + 4 \hat{g}^2 |z_2|^2) \over
p^2 (p^2 + \Sigma^2) (p^2 + 4\hat{g}^2|z_2|^2) (p^2 + \Sigma^2 + 4
\hat{g}^2 |z_2|^2)}
\end{eqnarray}
with respect to the cut-off $\Lambda$ to obtain the $\beta$ function:
\begin{eqnarray}
\beta(G) = \eta G - {\eta \over G_c} G^2 = \eta G - 4N \Sigma^2 C_d G^2
\end{eqnarray}
where we have substituted an explicit expression for the non trivial
ultraviolet fixed point (18):
\begin{eqnarray}
G_c = \frac{\eta}{4N \Sigma^2 C_d}, \;\;C_d = {2 \over \Gamma (d/2)
(4\pi)^{d/2}}.
\end {eqnarray}
Obviously, this fixed point goes to the origin in four
dimensions.

\subsection{Symmetries of the SUSY Gross-Neveu model and the mass spectrum
in the case B=0.}

We briefly discuss the symmetries of the lagrangian and compute the mass
spectrum in the  case B=0 for the sake of simplicity. Let us summarize
the symmetries of the problem:

$$
\begin{array}{lcc}
i) \: O(N) &  & \Phi_i \longrightarrow R_{ij} \Phi_j \cr
& & H_1 \longrightarrow H_1 \cr
& & H_2 \longrightarrow H_2
\end{array}
$$
$$
\begin{array}{lcc}
ii) \: U(1)_H &  & \Phi_i \longrightarrow e^{i \alpha/2} \Phi_i \cr
& & H_1 \longrightarrow e^{i \alpha} H_1 \cr
& & H_2 \longrightarrow e^{-i \alpha} H_2
\end{array}
$$
$$
\begin{array}{lcc}
iii) \: U(1)_R &  & \Phi_i \longrightarrow e^{i \frac{\beta}{2}}
\Phi_i (e^{-i \beta} \Theta) \cr
& & H_1 \longrightarrow e^{i \beta} H_1(e^{-i \beta} \Theta) \cr
& & H_2 \longrightarrow e^{i \beta} H_2 (e^{-i \beta} \Theta)
\end{array}
 $$

The $U(1)_R$ symmetry is broken by a $B \neq 0$ term.
The non-trivial vacuum has $<z_i> = 0 \: , \: <z_2> \neq 0 \: , \: <F_1>
\neq 0 \: , \: <z_1> = <F_2> = 0$. So, looking at the transformation laws
under the three symmetry groups indicates that:

- O(N) remains unbroken;

- $U(1)_H$ and $U(1)_R$ are separately broken, but the combination
$U(1)_{H + R}$ ($\beta = \alpha$) remains unbroken; on the other hand
the orthogonal combination $U(1)_{H - R}$ corresponding to $\beta = - \alpha$
is spontaneously broken and yields
a Goldstone boson in the mass spectrum.

To determine the full spectrum, we can use the constraint (9)
to eliminate $F_1$
from $V_{eff}$, obtained from (5) and (7):

\begin{eqnarray}
V_{eff} (z_1,z_2,F_2(z_1,z_2))  = m^2 |z_2|^2 + \Delta^2 |z_1|^2
- m(z_1 F_2 + z_1^* F_2^*) \cr
+ \frac{N}{2} \int \frac{{d^d}p}{(2 \pi)^d} [\ln (( p^2 + \Sigma^2
+ 4g^2 |{z_2}|^2)^2 - 4g^2 F_2 F_2^*) - 2 \ln (p^2 + 4g^2 |z_2|^2)]
\end{eqnarray}

The mass matrix for a function $V(x_i,F_k(x_j))$ is given by the
general formula (repeated indices are summed):

\begin{eqnarray}
\frac{d^2 V}{dx_i dx_j} = \frac{\partial^2 V}{\partial x_i \partial x_j}
+ \frac{\partial F_k}{\partial x_j} \frac{\partial^2 V}{\partial x_i
\partial F_k}
\end{eqnarray}
where we have used the constraint equation for $F_k$: $\partial V /
\partial F_k = 0$.

Making use of the explicit constraint equation for $F_2$:

\begin{eqnarray}
m z_1^* + 2 g^2 F_2 N \int \frac{{d^d}p}{(2 \pi)^d} \frac{1}{( p^2 + \Sigma^2
+ 4g^2 |{z_2}|^2)^2 - 4g^2 {F_2 F_2^*}} = 0
\end{eqnarray}

we obtain:
\vskip .2cm
\underline{i) in the trivial vacuum.}
\vskip .2cm

Here, the Goldstone boson disappears since no symmetry is broken. Indeed, the
mass matrix elements are:

\begin{eqnarray}
\frac{d^2 V}{dz^2_1} = \frac{d^2 V}{dz^{*2}_1}&=& \frac{d^2 V}{dz^2_2} =
\frac{d^2 V}{dz^{*2}_1} = 0     \cr
\frac{d^2 V}{dz^*_2 dz_2} = m^2 [ 1 - 4 N G &\Sigma^2 &
\int{{d^d p\over (2\pi)^d}\frac{1}{p^2
(p^2 + \Sigma^2 )^2}} ] = m^2 [ 1 - \frac {G}{G_c} ]    \cr
\frac{d^2 V}{dz^*_1 dz_1} &=& \Delta^2 + \frac{m^2}{2g^2 N
\int{{d^d p\over (2\pi)^d}\frac{1}{(p^2 + \Sigma^2 )^2}} }
\end{eqnarray}

So, for $G > G_c$, the trivial vacuum is unstable and the other vacuum
(see below) becomes stable.

\vskip .2cm
\underline{ii) in the nontrivial vacuum ($z_1 = F_2 = 0, z_2, F_1 \neq 0$).}
\vskip .2cm

$$
\frac{d^2 V}{dz_1 dz_2} = \frac{d^2 V}{dz^*_1 dz_2} = \frac{d^2 V}{dz^2_1} = 0
$$

The computation reveals two scalars of mass
$m_1$ (a scalar and a pseudoscalar) in the $H_1$ sector, the expected Goldstone
boson and a scalar of mass $m_2$ in the $H_2$ sector. Using the notation
$m_T \equiv \sqrt{4g^2 |z_2|^2}$, we write their masses as:

\begin{eqnarray}
m_1^2 & = & \Delta^2 + \frac{m^2}{2g^2 N \int{{d^d p\over (2\pi)^d}
\frac{1}{(p^2 + \Sigma^2 + m^2_T)^2}}} \cr
m_2^2 & = & 8 N g^2 m^2_T \int{ {d^d p\over (2\pi)^d}\left[
\frac{1}{(p^2 + m^2_T)^2} - \frac{1}{(p^2 + \Sigma^2 + m^2_T)^2}\right]}.
\end {eqnarray}

In presence of the gauge interactions, we will see that $m_2$ tends to
decrease and this solution may disappear for large gauge couplings.

\section{ The vacuum structure of the gauged model.}
\vskip .5cm

Until now, we have only studied, in the leading $\frac{1}{N}$ approximation,
global symmetries spontaneously broken  by fermionic condensates.
To make connection
with the standard model, we now include gauged
degrees of freedom and consider models of gauge symmetry
$SU(N) \otimes SU(2)_L \otimes U(1)_Y$ with global $SU(N)$ and local $SU(2)_L
\otimes U(1)_Y$. Because we are mainly concerned with the
scalar Higgs sector, we only take into account the electroweak gauge
sector leaving aside the $SU(N)$ color part which introduces technical
difficulties such as the handling of the planar graphs in the leading
approximation \cite{am}. In any case, the Higgs particles being color
singlets, the
$SU(N)$ part introduces corrections that will not change the vacuum
structure of the electroweak sector.

The interest in the gauged $SU(2)_L \times U(1)_Y$ model comes mainly
from the fact that, in the minimal supersymmetric standard model (MSSM)
the only quartic self interactions in the Higgs potential are given by the
corresponding gauge couplings which are thus essential to the vacuum
structure. Hence any
comparison of a supersymmetric condensate  model with the MSSM must include
them. We start with the following lagrangian:

\begin{eqnarray}
{\cal L}   = & &
\int {d^4}\Theta \: \{  [{\Phi_\alpha^+}e^{q_1YV_1+q_2V_2}{\Phi_\alpha} +
{\tilde{\Phi}_\alpha^+}e^{q_1 \tilde{Y}V_1}{\tilde{\Phi}_\alpha} ]
(1-{\Sigma^2}{\Theta^2}{{\bar \Theta}^2}) \cr
& & + H_1^+ e^{q_1Y_1V_1+q_2V_2} H_1
(1-{\Delta^2}{\Theta^2}{{\bar \Theta}^2}) \} \cr
& & +  \int {d^2} \Theta \:  [m {H_1} {H_2}(1 + B \Theta^2)
- g {H_2} {\Phi_\alpha}{\tilde{\Phi}_\alpha}] \cr
& & +\int {d^2} {\bar \Theta} \:  [m{H_1^+} {H_2^+} (1 + B {\bar \Theta}^2)
- g {H_2^+} {\Phi_\alpha^+}{{\tilde{\Phi}}_\alpha^+}]
\end{eqnarray}
where $\alpha = 1 \ldots N$ are color indices,  $q_1$ and
$q_2$ are the $U(1)_Y$ and the $SU(2)_L$ gauge couplings respectively.
Note that $H_1$, $H_2$, and $\Phi_\alpha$
are now $SU(2)$ doublets ($\Phi_\alpha = \Phi_{\alpha i},i=1,2$, etc. and
$H_1H_2 = \epsilon_{ij} H_{1i} H_{2j}$) whereas
$\tilde{\Phi}_\alpha$ are $SU(2)$ singlets.
Gauge invariance requires :

\begin{eqnarray}
Y_1 = - Y_2 =  Y + \tilde{Y}.
\end{eqnarray}

The D components for $SU(2)$ and $U(1)$ vector superfields are given
respectively by:

\begin{eqnarray}
D^a = q_2 ( z^+_\alpha T^a z_\alpha  + z^+_1 T^a z_1 ) \cr
D = q_1 ( Y z^+_\alpha z_\alpha
+ \tilde{Y} \tilde{z}^+_\alpha \tilde{z}_\alpha + Y_1 z_1^+ z_1 )
\end{eqnarray}
with $T^a$ being a representation of the $SU(2)$ generators.
Writing the component
lagrangian and integrating over:

\begin{eqnarray}
F^*_{\alpha j} = g \epsilon_{i j} z_{2 i} \tilde{z}_\alpha \cr
\tilde{F}^*_\alpha = g \epsilon_{i j} z_{2 i} z_{\alpha j},
\end{eqnarray}
we obtain:

\begin{eqnarray}
{\cal L} & = &  {F^+_1}{F_1} + { z^+_1}(\Box - \Delta^2){z_1}
- i \bar{\Psi}_{1i} \bar{\sigma}^m \partial_m \Psi _{1i}
+ {z^+_\alpha} ( \Box - \Sigma^2) {z_\alpha} \cr
& & - i\bar{\Psi}_{\alpha i} \bar{\sigma}^m \partial_m \Psi_{\alpha i}
+ \tilde{z}^*_\alpha ( \Box - \Sigma^2) \tilde{z}_\alpha
- i\bar{\tilde{\Psi}}_\alpha
\bar{\sigma}^m \partial_m \tilde{\Psi}_\alpha \cr
& & + m \epsilon_{ij} (z_{1i} F_{2j} + z_{2j} F_{1i} - \Psi_{1i} \Psi_{2j} +
B z_{1i} z_{2j} + h.c.) \cr
& & - g \epsilon_{ij} (F_{2i}z_{\alpha j} \tilde{z}_\alpha
- z_{2i}\Psi_{\alpha j}
\tilde{\Psi}_\alpha - \tilde{z}_\alpha \Psi_{2i} \Psi_{\alpha j} - z_{\alpha j}
\Psi_{2i} \tilde{\Psi}_\alpha + h.c.) \cr
& & - \frac{q^2_2 + 4 q_1^2 ( Y +\tilde{Y} )^2}{8} (z_1^+ z_1)^2 \cr
& & - z^*_{\alpha i} \{  \delta_{ij} [ g^2 z_2^+ z_2
- Q^2 z^+_1 z_1 ] + \frac{q_2^2}{2} z_{1i} z^*_{1j}
- g^2 z_{2i} z^*_{2j}  \} z_{\alpha j} \cr
& & - |\tilde{z}_\alpha|^2 [ g^2 z^+_2 z_2
+ \tilde{Q}^2  z^+_1 z_1 ]
+ {\cal O} ( z^4_\alpha )
\end{eqnarray}
where
\begin{eqnarray}
Q^2 = \frac{q_2^2}{4} -q_1^2 Y(Y+ \tilde{Y})  \cr
\tilde{Q}^2 = q_1^2 \tilde{Y}(Y+ \tilde{Y})
\end{eqnarray}
(we will see below that $Q^2,\tilde{Q}^2$ are positive).
The last term ${\cal O} ( z^4_\alpha )$ in (46)
plays no role in the computation of the
effective potential in the $H_1$, $H_2$ Higgs sector to the leading order
in $N$. This potential reads in $d=4$ spacetime dimensions (cf. (6))

\begin{eqnarray}
V_{eff} = V_{tree} + {1 \over 32 \pi^2} \left\{
\Lambda^2 {\cal {ST}}r M^2 + {1\over2} {\cal {ST}}r (M^4 \ln{M^2 \over
\Lambda^2}) - {1\over 4} {\cal {ST}}r M^4 + {\cal
O}\left({M^2\over\Lambda^2}\right) \right\}
\end{eqnarray}

We wish to compute $V_{eff}$ in the configuration:

\begin{eqnarray}
z_1 = \left(
  \begin{array}{c}
   h_1 \\ 0
  \end{array}
\right) ,
z_2 = \left(
  \begin{array}{c}
   0 \\ h_2
  \end{array}
\right) ,
F_1 = \left(
  \begin{array}{c}
   f_1 \\ 0
  \end{array}
\right) ,
F_2 = \left(
  \begin{array}{c}
   0 \\ f_2
  \end{array}
\right)
\end{eqnarray}
which does not break  charge conservation. The eigenvalues of the
scalar mass matrix are:

\begin{eqnarray}
m_1^2 = \Sigma^2 - Q^2 |h_1|^2 \nonumber
\end{eqnarray}
\begin{eqnarray}
m^2_{2,3} & = & \Sigma^2 + g^2 |h_2|^2 + \frac{1}{2} ( -Q^2 + \tilde{Q}^2 +
\frac{q_2^2}{2} ) |h_1|^2 \cr
& & \pm \sqrt{\frac{1}{4} ( -Q^2 + \tilde{Q}^2 +
\frac{q_2^2}{2})^2 |h_1|^4 + g^2 |f_2|^2}
\end{eqnarray}
whereas for the fermions:

\begin{eqnarray}
(m^+_F m_F)_{\alpha_i \alpha_j} = g^2 (z^+_2 z_2 \delta_{ij} - z^+_{2j} z_{2i})
\nonumber
\end{eqnarray}
\begin{eqnarray}
(m^+_F m_F)_{\tilde{\alpha} \tilde{\alpha}} = g^2 |h_2|^2
\end{eqnarray}

Then:

\begin{eqnarray}
\frac{1}{N} {\cal{ST}}r M^2 = 6\Sigma^2 + 2 (-2Q^2 + \tilde{Q}^2
+ \frac{q_2^2}{2})|h_1|^2
\end{eqnarray}
and to avoid a field dependent quadratic divergence we set:

\begin{eqnarray}
-2Q^2 + \tilde{Q}^2 + \frac{q_2^2}{2} =
q_1^2 (2Y + \tilde{Y})(Y + \tilde{Y}) = 0
\end{eqnarray}
which reduces to $2Y + \tilde{Y} = 0$ and is equivalent to the usual
condition of the cancellation of gauge anomalies in the standard model.

Using this condition we obtain from (47) that:
\begin{eqnarray}
Q^2 &=& {q_2^2 \over 2} + q_1^2 (Y + \tilde{Y})^2 \geq {q_2^2 \over 2}, \cr
\tilde{Q}^2 &=& 2 q_1^2 (Y + \tilde{Y})^2.
\end {eqnarray}
Using also:
\begin{eqnarray}
\frac{1}{N} {\cal{ST}}r M^4 &=& 6\Sigma^4
+ 4 [(Q^2 - \tilde{Q})^2)^2 +  Q^2 \tilde{Q}^2]
|h_1|^4 \cr
&+&4g^2 (-Q^2 + \tilde{Q}^2 + \frac{q_2^2}{2}) |h_1|^2 |h_2|^2 \cr
&+&4 \Sigma^2 (-2Q^2 + \tilde{Q}^2 + \frac{q_2^2}{2}) |h_1|^2
+ 8 \Sigma^2 g^2 |h_2|^2 + 4g^2 |f_2|^2
\end{eqnarray}
we obtain the one-loop (leading $\frac{1}{N}$) effective potential:
\footnote{Some details on the computation are given in section i) of the
appendix.}
\begin{eqnarray}
V_{eff} & = & - |f_1|^2 + \Delta^2 |h_1|^2 - m (h_1 f_2 + h_2 f_1
+ B h_1 h_2 + h.c. )
+ \frac{Q^2}{2} |h_1|^4 \cr
& & + \frac{N}{32\pi^2} \{ 6\Sigma^2 \Lambda^2 + \sum_{i=1}^{3} m_i^4
\ln{\frac{m_i^2}{\Lambda^2}} - 2g^4 |h_2|^4 \ln{\frac{g^2 |h_2|^2}{\Lambda^2}}
\cr
& & -\frac{3}{2} \Sigma^4 - [(Q^2 - \tilde{Q}^2)^2 +  Q^2 \tilde{Q}^2]
|h_1|^4 - g^2 Q^2 |h_1|^2 |h_2|^2 \cr
& & - 2\Sigma^2 g^2 |h_2|^2 -g^2 |f_2|^2 \} + {\cal O} (\frac{M^2}{\Lambda^2}).
\end{eqnarray}

When studying the correspondence with the minimal supersymmetric standard
model (MSSM), it is particularly illuminating to consider the limit $B
\rightarrow 0$. Indeed when $B=0$, it is well-known that the MSSM has a
richer vacuum structure since, depending on the parameters, besides the
trivial minimum $h_1 = h_2 =0$ and the completely nontrivial minimum, we
have the mixed possibilities $h_1=0, h_2 \neq 0$ (and $h_2=0, h_1 \neq 0$).
It is interesting in its own sake to see if such a vacuum structure is also
present in the model considered here. On the other hand, the
phenomenologically relevant case corresponds to nonzero $B$ soft terms.

Let us consider first the case $B=0$. The saddle point equations
corresponding to the potential $V_{eff}$ in (56) are given by:

\begin{eqnarray}
f_1 = - m h_2^* \nonumber
\end{eqnarray}
\begin{eqnarray}
m f^*_2 = h_1 \left\{ \Delta^2 + Q^2 |h_1|^2 + \frac{N}{32\pi^2}
\left[ -Q^2 (2m_1^2 \ln
{\frac{m_1^2}{\Lambda^2}} - m_2^2 \ln{\frac{m_2^2}{\Lambda^2}}
- m_3^2 \ln{\frac{m_3^2}{\Lambda^2}} ) \right. \right. \cr
\left. \left. + (Q^2 + \tilde{Q}^2 - \frac{q_2^2}{2})^2
|h_1|^2 \frac{m_2^2 \ln{\frac{m_2^2} {\Lambda^2}} - m_3^2
\ln{\frac{m_3^2}{\Lambda^2}}}{m_2^2 - m_3^2} \right] \right\}
\nonumber
\end{eqnarray}
\begin{eqnarray}
m h_1^* = \frac{Ng^2}{16\pi^2} f_2 \frac{m_2^2 \ln{\frac{m_2^2}
{\Lambda^2}} - m_3^2 \ln{\frac{m_3^2}{\Lambda^2}}}{m_2^2 - m_3^2} \nonumber
\end{eqnarray}
\begin{eqnarray}
m f_1^* = \frac{Ng^2}{16\pi^2} h_2 \left( m_2^2 \ln{\frac{m_2^2}{\Lambda^2}}
+ m_3^2 \ln{\frac{m_3^2}{\Lambda^2}} - 2g^2 |h_2|^2
\ln{\frac{g^2 |h_2|^2}{\Lambda^2}} \right)
\end{eqnarray}
Combining them, one obtains the following two equations:
\begin{eqnarray}
0 & = & f_2 \left\{ m^2 - \frac{Ng^2}{16\pi^2} \frac{m_2^2 \ln{\frac{m_2^2}
{\Lambda^2}} - m_3^2 \ln{\frac{m_3^2}{\Lambda^2}}}{m_2^2 - m_3^2}  \left(
\Delta^2 + Q^2 |h_1|^2 \right.\right. \cr
& & + \frac{N}{32\pi^2} [ -Q^2 (2m_1^2 \ln{\frac{m_1^2}{\Lambda^2}} - m_2^2
\ln{\frac{m_2^2}{\Lambda^2}} - m_3^2 \ln{\frac{m_3^2}{\Lambda^2}} ) \cr
& & \left.\left.+ (Q^2 + \tilde{Q}^2 - \frac{q_2^2}{2})^2
|h_1|^2 \frac{m_2^2
\ln{\frac{m_2^2}{\Lambda^2}} - m_3^2 \ln{\frac{m_3^2}{\Lambda^2}}}{m_2^2
- m_3^2} ] \right) \right\}
\end{eqnarray}
\begin{eqnarray}
0 = h_2 \left\{ m^2 + \frac{Ng^2}{16\pi^2} \left( m_2^2
\ln{\frac{m_2^2}{\Lambda^2}}
+ m_3^2 \ln{\frac{m_3^2}{\Lambda^2}} - 2g^2 |h_2|^2 \ln{\frac{g^2 |h_2|^2}
{\Lambda^2}} \right) \right\}
\end{eqnarray}
where $m_i^2$ are given in (50) and $Q,\tilde{Q}$ in (47) or (54).

This system has possibly four different solutions:

  i) The trivial one : $h_1 = h_2 = f_1 = f_2 = 0$.

 ii) The solution $f_2 = h_1 = 0, h_2 \neq 0$ which is independent
of the gauge couplings and gives through (59) the usual gap equation (20)
(up to a factor 2 which comes from the different field content):
\begin{eqnarray}
G^{-1}={N \over 8\pi^2}
\left[\Sigma^2\ln{\frac{\Lambda^2}{\Sigma^2 + g^2 |h_2|^2}}
- g^2 |h_2|^2 \ln{\frac{\Sigma^2 + g^2 |h_2|^2}{g^2|h_2|^2}} \right].
\end{eqnarray}
This is the vacuum which was studied by Clark, Love and Bardeen \cite{ai} and
we will sometimes refer to it as the CLB vacuum.

iii) The solution $f_1 = h_2 = 0$ and $h_1 \neq 0$ given by (58).

 iv) The completely nontrivial solution
$h_1 \neq 0$ , $h_2 \neq 0$ , $f_1 \neq 0$ and $f_2 \neq 0$.

We thus recover the four different vacua present in the MSSM.
The last two extrema are difficult to study analytically.

For the trivial extremum, the mass matrix computed in section 4.2
remains valid, up to trivial factors of 2. So, for $G > G_c$ with $G_c$
given by an expression similar to (18), namely
\begin{eqnarray}
G_c = \frac{8\pi^2}{N \Sigma^2\ln{\frac{\Lambda^2}{\Sigma^2}}}.
\end{eqnarray}
it becomes unstable.

In the extremum ii), the CLB vacuum, the only difference compared with the
analysis of section 4.2 lies in the mass matrix element:
\begin{eqnarray}
\frac{d^2 V}{dh^*_1 dh_1} & = & \Delta^2
+ \frac{m^2}{ N g^2 \int{{d^d p\over (2\pi)^d}
\frac{1}{(p^2 + \Sigma^2 + g^2 |h_2|^2)^2}}} \cr
&  & - N Q^2 g^2 |h_2|^2 \int {d^d p\over (2\pi)^d}
{1 \over (p^2 + \Sigma^2)(p^2 + \Sigma^2 + g^2 |h_2|^2)},\nonumber
\end{eqnarray}
which reads in 4 dimensions
\begin{eqnarray}
\frac{d^2 V}{dh^*_1 dh_1} & = & \Delta^2 + [\frac{NG}{16\pi^2} \ln{\frac
{\Lambda^2}{\Sigma^2 + g^2 |h_2|^2}} - 1 ]^{-1} \cr
& & - \frac{N Q^2}{16\pi^2} [g^2 |h_2|^2 \ln{\frac{\Lambda^2}{\Sigma^2
+ g^2 |h_2|^2}} - \Sigma^2 \ln{\frac{\Sigma^2 + g^2 |h_2|^2}{\Lambda^2}} ]
\end{eqnarray}
For every given $G$, we have a value of Q beyond which this extremum is
unstable and cannot be the true vacuum: the gap equation
and the spectrum are then completely different. For example, we will
see below that there is a smooth transition to the extremum iv) which
has  an extra Goldstone boson.

Indeed, let us rediscuss the issue of the symmetries for the case at hand. The
symmetry of our lagrangian is $SU(N) \times SU(2)_L \times U(1)_Y
\times U(1)_R$. If we are interested in the neutral Higgs sector, we can
restrict ourselves to the transformations associated with the quantum
numbers $T_3$ , $Y$ and $R$. The charge $Q = T_3 + \frac{Y}{2}$
being conserved, we are left with the independent combinations
$T_3 -{Y\over 2} \pm R$. Under $T_3 -
\frac{Y}{2} + R$ the fields transform according to:

\begin{eqnarray}
h_1 \longrightarrow e^{i\alpha} h_1 & , & f_1 \longrightarrow f_1 \cr
h_2 \longrightarrow h_2 & , & f_2 \longrightarrow e^{-i\alpha} f_2
\end{eqnarray}
And under $T_3 - \frac{Y}{2} - R$ as:
\begin{eqnarray}
h_1 \longrightarrow h_1 & , & f_1 \longrightarrow e^{-i\beta} f_1 \cr
h_2 \longrightarrow e^{i\beta} h_2 & , & f_2 \longrightarrow f_2
\end{eqnarray}

In the case ii), $<h_2> \neq 0$ , $<h_1> = 0$, so the $T_3 - \frac{Y}{2}
- R$ symmetry is spontaneously broken, giving a Goldstone boson in the
neutral sector. In the third extremum iii), the broken symmetry is
$T_3 - \frac{Y}{2} + R$
being accompanied by its corresponding Goldstone boson. Both this two
symmetries are broken in the fourth, completely non trivial, extremum
leading to two massless bosons. If we restore a nonzero value for the soft
supersymmetry-breaking parameter $B$, $U(1)_R$ is no longer a symmetry
of the Lagrangian; all the last three extrema break $T_3 - \frac{Y}{2}$
and yield one Goldstone boson.

To study the smooth transitions between the four extrema,
we use the same technics as in section 3.
We stress that this scenario is only one of the possibilities; the complete
set of transitions (including first order) are obtained by comparing
the values of the energy for the four extrema as fonctions of the
couplings. We will however consider here only possible second order phase
transitions, in the spirit of the original work of Nambu and
Jona-Lasinio \cite{ad}.

Suppose that we start from the trivial minimum and that for
$G = G_c $ the system undergoes a second order phase transition from
this trivial extremum to one of the three nontrivial minima. Thus for
$G=G_c(1+\epsilon)$, $\epsilon \ll 1$, we have:
$f_1/\Sigma^2$ , $g^2 |h_2|^2/\Sigma^2$ ,
$4g^2 |f_2|^2/\Sigma^4$ , $|h_1|^2/\Sigma^2 \ll 1$.
We keep only the leading terms in the saddle point equations (57):

\begin{eqnarray}
f_1 & = & - m h_2^* \cr
m f_2^* & = & \Delta^2 h_1 \cr
h_1^* & = & - \frac{NmG_c}{16\pi^2} (\ln{\frac{\Lambda^2}{\Sigma^2}} -1) f_2
\cr
f_1^* & = & - \frac{NmG_c}{8\pi^2} \Sigma^2\ln
\left({\frac{\Lambda^2}{\Sigma^2}}\right) h_2
\end{eqnarray}
which, combined, give $h_1 = 0$ and (61).
Thus the only allowed second order transition from the trivial vacuum is to
the CLB vacuum ii) ($h_1 = 0, h_2 \neq 0$).
Indeed one recovers the standard value (18) for the critical coupling.
Note in particular that this critical value is independent of the gauge
couplings $Q$ and $\tilde{Q}$.

The next step is to study the smooth transitions at $G \sim \tilde{G}_c$
from the vacuum ii) to any of the other nontrivial vacua, which for that
matter can only be iv). At $G=\tilde{G}_c(1 + \epsilon)$, we have $|h_2|$
close to the value given by the gap equation (60) (in the following we note
$|h_2|_c$ the corresponding value):
$|h_2|^2 = |h_2|_c^2 + \delta |h_2|^2, g^2 \delta|h_2|^2/\Sigma^2 \ll 1$.
On the other hand, we have as before $|h_1|^2/\Sigma^2,
4g^2|f_2|^2/\Sigma^4 \ll 1$. The leading terms now read:
\begin{eqnarray}
f_1 & = & - m h_{2c}^* \cr
m f_2^* & = & \left\{\Delta^2  - \frac{NQ^2}{16\pi^2} (g^2 |h_2|_c^2 \ln{ \frac
{\Lambda^2}{\Sigma^2 + g^2 |h_2|_c^2}} + \Sigma^2 \ln{ \frac{\Sigma^2}
{\Sigma^2 + g^2 |h_2|_c^2}} )\right\} h_1 \cr
h_1^* & = & - \frac{Nm\tilde{G}_c}{16\pi^2}
\left(\ln{\frac{\Lambda^2}{\Sigma^2 +
g^2 |h_2|^2_c}} -1\right) f_2 \cr
f_1^* & = & - \frac{Nm\tilde{G}_c}{8\pi^2} \left(
\Sigma^2 \ln{\frac{\Lambda^2}{\Sigma^2 + g^2 |h_2|_c^2}}
- g^2 |h_2|_c^2 \ln{\frac{\Sigma^2 + g^2 |h_2|_c^2}{g^2|h_2|_c^2}} \right)
 h_{2c}
\end{eqnarray}
These equations can be recast into (60) (with $G=\tilde{G}_c$, $|h_2|^2
= |h_2|_c^2$), as expected since we consider a smooth transition from vacuum
ii) and
\begin{eqnarray}
1& =& \frac{N\tilde{G}_c}{16\pi^2} \left( \ln{\frac{\Lambda^2}{\Sigma^2+
g^2 |h_2|^2}} -1 \right) \cr
& & \left\{ \frac{NQ^2}{16\pi^2} \left[ g^2 |h_2|_c^2
\ln{\frac{\Lambda^2}{\Sigma^2 + g^2 |h_2|_c^2}}
- \Sigma^2  \ln{\frac{\Sigma^2 + g^2 |h_2|_c^2}{\Sigma^2}} \right]
- \Delta^2 \right\}
\end{eqnarray}
Details on the derivation are given in the appendix, section ii).

Equations (60) and (67) can be turned into an equation giving $Q^2$ in
terms of $|h_2|^2$ (which is itself a monotonous function of $G$ through
(60)). It can be  checked that $Q^2$ is itself a monotonous function of
$\tilde{G}_c$ on the interval $(\tilde{G}_c,+\infty)$, decreasing from
$+\infty$ to $0$. Inverting this function, we therefore obtain the critical
line between the standard vacuum ii) and the completely nontrivial vacuum
shown in Fig.2.

\begin{figure}[p]
\vspace{10cm}
\caption{Phase diagram in the ($G$,$Q^2$) plane for the gauged model.}
\end{figure}

Indeed, Fig.2 summarizes the critical values of the coupling $G$ as a
function of $Q$ for the smooth transitions (second order) between the
different vacua.

It is now easy to discuss the case $B \neq 0$ which is more relevant for
phenomenology but less rich in its vacuum structure. Only the trivial
($h_1=h_2=0$) and completely nontrivial
($h_1\neq 0, h_2 \neq 0$) configurations are allowed minima and the
critical line between the two minima is given by $G_c$ in (61),
irrespective of $Q$. One indeed checks from the scalar mass matrix that the
trivial vacuum is unstable for $G>G_c$.

\section{Conclusions.}

  We have studied the vacuum structure in supersymmetric models with scalar
condensates by using the effective potential approach in the leading
$\frac{1}{N}$ approximation. The resulting saddle point equations were analyzed
and the phase transitions betwen the different extrema obtained by
linearization
around the critical surfaces.

  We have closely studied the equivalence of the supersymmetric top-antitop
condensate model with the MSSM. In the case where the soft term known as B is
zero\footnote{This case is chosen because of the richer structure of the
vacuum.},
the vacuum structure of the condensate model is the same as the one in the MSSM
at tree level, {\sl{provided}} we take into account the $SU(2)_L \times U(1)_Y$
couplings which have leading $\frac{1}{N}$ contributions.
We find four extrema; the phase diagram now involves the gauge couplings as
well
as the original four-fermion coupling G.

 In the case $B \neq 0$, the vacuum solutions still depend on the
gauge couplings.
But the critical surface which separates the trivial vacuum from the
non-trivial one is determined only by the self-coupling G in distinction with
the case $B=0$. This critical surface is therefore identical to the one
obtained when electroweak gauge interactions are turned off.
In this version of the model, we have studied in detail the role played by
the soft supersymmetry breaking terms, in particular $B$, on the breaking
of chiral symmetry.

 The methods developped in this paper are readily applicable to other studies
such as temperature dependence of the condensate solution or similar analyses
of
dynamical symmetry breaking in non-minimal supersymmetric models.
\subsection*{   Acknowledgments}
We wish to thank C.Savoy for many valuable discussions and C.Teodorescu
for help in the numerical analyses.

\newpage
\section*{Appendix.}

\vskip .5cm

  {\bf i) The effective potential in the gauged model.}

\vskip .3cm
We can write the lagrangian in (46) as
\begin{eqnarray}
{\cal L} & = &  {F^+_1}{F_1} + { z^+_1}(\Box - {\Delta^2}){z_1}
-i{{\bar \Psi}_{1i}}{{\bar \sigma}^m}{\partial _m} \Psi_{1i}
- i\bar{\Psi}_{\alpha i}{{\bar \sigma}^m}{\partial_m}{\Psi}_{\alpha i} \cr
& & - i\bar{\tilde{\Psi}}_{\alpha }{{\bar \sigma}^m}{\partial_m}\tilde
{\Psi}_{\alpha } + m \epsilon_{ij}(z_{1i} F_{2j}+z_{2j}F_{1i}
- \Psi_{1i}\Psi_{2j} + h.c.) \cr
& & -\frac{q_2^2+4q_1^2(Y+\tilde{Y})}{8}(z_1^+z_1)^2
+ Bm(\epsilon_{ij} z_{1i}z_{2j}+h.c.)\cr
& & -g\epsilon_{ij} (z_{2i}\Psi_{\alpha j} \tilde{\Psi}_{\alpha}
- 2{z_i}{\Psi_j}{\Psi_2} + h.c.)\cr
& & -(z^*_{\alpha i} \tilde{z}_{\alpha})
\left( \begin{array}{cc}
       X\delta_{ij}+\frac{q_2^2}{2}z_{1i}z^*_{1j}-g^2z_{2i}z^*_{2j}
& g\epsilon_{ki}F_{2k} \\
       g\epsilon_{kj}F_{2k} & \tilde{X}
\end{array} \right)
\left( \begin{array}{c} z_{\alpha j} \\ \tilde{z}^*_{\alpha} \end{array}
\right)
\end{eqnarray}
with the notations:
\begin{eqnarray}
X = -\Box + g^2 z^+_2 z_2 + ( q_1^2 Y(Y+ \tilde{Y}) - \frac{q_2^2}{4} )
z^+_1 z_1 + \Sigma^2, \cr
\tilde{X} =   -\Box + g^2 z^+_2 z_2 + q_1^2 Y(Y+ \tilde{Y}) z^+_1 z_1
+ \Sigma^2.
\nonumber
\end{eqnarray}
If M and $m_f$ are the scalar and the fermion mass matrices, then:
\begin{eqnarray}
\ln{det(p^2+M^2)} & = & N\ln \{ (X\tilde{X} -g^2F^+_2F_2)
(X+\frac{q^2_2}{2}z^+_2z_2
-g^2z^+_2z_2) \cr
& &-g^4(z^+_2z_2F^+_2F_2-z^+_2F_2F^+_2z_2) \cr
& & -\frac{q^2_2}{2}g^2\tilde{X}(z^+_1z_1z^+_2z_2-z^+_1z_2z^+_2z_1) \cr
& & +\frac{q^2_2}{2}g^2(z^+_1z_1F^+_2F_2-z^+_1F_2F^+_2z_1) \}
\end{eqnarray}
where from now on we replace $\Box$ by $-p^2$ in $X$ and $\tilde{X}$ above.
Then
\begin{eqnarray}
\ln{det(p^2+m_f^2)} & & = N\ln{det
\left( \begin{array}{cc}
       (p^2+g^2z^+_{2j}z_{2i})\delta_{ij} -g^2z^+_{2j}z_{2i} & 0 \\
       0 & p^2+g^2z^+_2z_2
       \end{array}
\right) } \cr
& & = N\ln{p^2(p^2+g^2z^+_2z_2)}
\end{eqnarray}
Finally,
\begin{eqnarray}
V_{eff} & = & - F^+_1 F_1 + \Delta^2 z^+_1 z_1
- m \epsilon_{ij} (z_{1i} F_{2j} + z_{2j} F_{1i} + h.c. ) \cr
& & + \frac{q^2_2 + 4 q_1^2 ( Y +\tilde{Y} )^2}{8} (z_1^+ z_1)^2 \cr
& & + N \int \frac{{d^4}p}{(2 \pi)^4} \: [ \: \ln \{ (X \tilde{X}
- g^2 F_2^+ F_2) ( X+ \frac{q_2^2}{2} z^+_1 z_1 - g^2 z_2^+ z_2 ) \cr
& & - \frac{q_2^2}{2} g^2\tilde{X} (z^+_1 z_1 z^+_2 z_2
- z^+_1 z_2 z^+_2 z_1 ) \cr
& & + \frac{q_2^2}{2} g^2 (z^+_1 z_1 F^+_2 F_2 - z^+_1 F_2 F^+_2 z_1 )
- g^4 ( z_2^+ z_2 F_2^+ F_2 - z_2^+ F_2 F_2^+ z_2 ) \}  \cr
& & - \ln \{ p^2 (p^2 + g^2 z_2^+ z_2)^2 \} \: ]
\end{eqnarray}
which by integration in the configuration (49) gives the result written as
equation (56).

\vskip .5cm
  {\bf ii) The phase transition from the Clark-Love-Bardeen (CLB) vacuum to the
completely nontrivial one (assuming $B=0$).}
\vskip .3cm

    We have verified in equation (65) that a second order phase
transition from the trivial to the CLB vacuum ii) occurs at $G = G_c$.
We now verify that for $G \gg G_c$, one encounters a new smooth transition
to the completely nontrivial minimum. Following the method developped in
section 3, we determine the corresponding critical surface $f(G,q_1,q_2) = 0$
by requiring that, when we are in the vicinity of this critical surface, {\it
i.e.} $(f + \epsilon g)(G,q_1,q_2) = 0, \epsilon \ll 1$,
we have $\frac{|h_1|^2}{\Sigma^2}$ , $4\frac{g^2 |f_2|^2}{\Sigma^4} \ll 1$ and
$g^2 |h_2|^2 = g^2 |h_2|^2_c (1 + \delta x)$ where $g^2 |h_2|^2_c$ is the
CLB solution and $\delta x \ll 1$. Fixing $G = G_0$ we search
for a $Q_0^2$ beyond
which we have a completely nontrivial solution ($h_1 \neq 0, h_2 \neq 0$).
Writing $Q^2 = Q_0^2 (1 + \epsilon)$ and making a limited Taylor series
expansion in the saddle point equations (57), we obtain the following system of
equations:
\begin{eqnarray}
f_1 & = & - m h_2^* \cr
m f_2^* & = & \{ A_1 + B_1 |h_1|^2 \} h_1 \cr
m h_1^* & = & \frac{Ng_c^2}{16\pi^2} \{ A_2 + B_2 |h_1|^2 \} f_2
\cr
m f_1^* & = & \frac{Ng_c^2}{16\pi^2} \{ A_3 + B_3 |h_1|^2 \} h_2
\end{eqnarray}
where the functions $A_i$ and $B_i$ are of the form (in the following,
$|h_2|^2$
means the CLB solution $|h_2|^2_c$ and $G$ and $Q^2$ are $G_0$ and $Q_0^2$):

\begin{eqnarray}
A_1 & = & \Delta^2 - \frac{NQ^2(1+\epsilon)}{16\pi^2} (g^2 |h_2|^2 \ln{ \frac
{\Lambda^2}{\Sigma^2 + g^2 |h_2|^2}} + \Sigma^2 \ln{ \frac{\Sigma^2}
{\Sigma^2 + g^2 |h_2|^2}} ) \cr
& & - \frac{NQ^2}{16\pi^2} g^2 |h_2|^2 \ln{ \frac{\tilde{\Lambda}^2}{\Sigma^2
+ g^2 |h_2|^2}} \delta x \nonumber
\end{eqnarray}
\begin{eqnarray}
A_2 = 1 - \ln{\frac{\Lambda^2}{\Sigma^2 + g^2 |h_2|^2}}
+ \frac{g^2|h_2|^2}{\Sigma^2 + g^2 |h_2|^2} \delta x
\nonumber
\end{eqnarray}
\begin{eqnarray}
A_3 & = & - 2\Sigma^2 \ln{\frac{\Lambda^2}{\Sigma^2 + g^2 |h_2|^2}}
+ 2 g^2 |h_2|^2 \ln{\frac{\Sigma^2 + g^2 |h_2|^2}{g^2|h_2|^2}} (1 + \delta x)
\nonumber
\end{eqnarray}
\begin{eqnarray}
B_1 & = & Q^2 - \frac{N}{32\pi^2} [Q^2 (3 - 2\ln{\frac{\Lambda^2}{\Sigma^2}}
- \ln{\frac{\Lambda^2}{\Sigma^2 + g^2 |h_2|^2}}) \cr
& & + (Q^2 + \tilde{Q^2} - \frac{q_2^2}{2})^2
\ln{\frac{\tilde{\Lambda}^2}{\Sigma^2
+ g^2 |h_2|^2}}] \nonumber
\end{eqnarray}
\begin{eqnarray}
B_2 = \frac{Q^2}{\Sigma^2 + g^2 |h_2|^2} \nonumber
\end{eqnarray}
\begin{eqnarray}
B_3 = - Q^2 \ln{\frac{\tilde{\Lambda}^2}{\Sigma^2 + g^2 |h_2|^2}} \nonumber
\end{eqnarray}
where $\tilde{\Lambda}^2 = \Lambda^2/e$.

To find the critical surface, only the ${\cal O}(1)$ terms are kept in
equations (72). They are easily combined to give equation (67).
The critical surface indeed coincides with the instability surface for the
CLB vacuum where the matrix element ${\frac{d^2V}{dh_1^*dh_1}}_{h_1=0}$
becomes negative. This observation is a strong support for the
above mentioned scenario: the completely nontrivial vacuum occurs for values of
the couplings at which the CLB vacuum becomes unstable. To verify
that the scenario is really possible, we compute $\frac{|h_1|^2}{\Sigma^2}$,
$\frac{g^2 |h_2|^2}{\Sigma^4}$ and $\delta x$ as functions of $\epsilon$.
Decomposing $A_i = A_{i0} + A_i^\prime$, where $A_i^\prime$ is a collection of
small terms, we obtain the equations:

\begin{eqnarray}
2g^2 |h_2|^2 \ln{\frac{\Sigma^2 + g^2 |h_2|^2}{g^2|h_2|^2}} \delta x
+ B_3 |h_1|^2 & = & 0 \cr
A_1^\prime A_{20} + A_{10} A_2^\prime + (A_{10} B_2 + A_{20} B_1) |h_1|^2 &
= & 0
\end{eqnarray}

Because $B_3 < 0$ we find $\delta x$ positive and proportional to $|h_1|^2$ and
$|h_1|^2 = C \epsilon$ where C is a rather lengthy expression:
\begin{eqnarray}
C & = & \frac{NQ^2}{16\pi^2} \ln{\frac{\tilde{\Lambda}^2}{\Sigma^2
+ g^2 |h_2|^2}} [\ln{\frac{\Sigma^2 + g^2 |h_2|^2}{g^2|h_2|^2}}]^{-1} \cr
& & + Q^2 ln{\frac{\tilde{\Lambda}^2}{\Sigma^2 + g^2 |h_2|^2}}
[(\Sigma^2 + g^2 |h_2|^2) \ln{\frac{\Sigma^2 + g^2 |h_2|^2}{g^2|h_2|^2}}]^{-1}
\cr
& & (g^2 |h_2|^2 \ln{\frac{\Lambda^2}{\Sigma^2 + g^2 |h_2|^2 }} -\Sigma^2
 \ln{\frac{\Sigma^2 + g^2 |h_2|^2}{g^2|h_2|^2}} ) \cr
& & + \frac{2Q^2}{\Sigma^2 + g^2 |h_2|^2}
(g^2 |h_2|^2 \ln{\frac{\Lambda^2}{\Sigma^2 + g^2 |h_2|^2 }} -\Sigma^2
 \ln{\frac{\Sigma^2 + g^2 |h_2|^2}{g^2|h_2|^2}} ) \cr
& & - \ln{\frac{\tilde{\Lambda}^2}{\Sigma^2 + g^2 |h_2|^2}}
\{ Q^2 + \frac{N}{32\pi^2} [Q^2 (-3 + 2\ln{\frac{\Lambda^2}{\Sigma^2}}
+ \ln{\frac{\Lambda^2}{\Sigma^2 + g^2 |h_2|^2}}) \cr
& & - (Q^2 + \tilde{Q^2} - \frac{q_2^2}{2})^2
\ln{\frac{\tilde{\Lambda}^2}{\Sigma^2
+ g^2 |h_2|^2}}] \} >0
\nonumber
\end{eqnarray}
Henceforth, for $\epsilon > 0$, we have $|h_1|^2 > 0$ , $\delta x > 0$ and for
$\epsilon < 0$, $|h_1|^2 = \delta x = 0$ which corresponds to a usual second
order phase transition.

\vskip .5cm
  {\bf iii) The gap equations for the non-gauged, $B \neq 0$ case.}
\vskip .3cm
The solutions of equations (16) may be rewritten in the following form:
\begin{eqnarray}
G^{-1} & = & \frac{N}{8\pi^2} [ (\Sigma^2 + 4g^2 |{z_2}|^2 - \frac{BF_2}{2z_2}
)
\ln{ \frac{\Lambda^4}{(\Sigma^2 + 4g^2 |{z_2}|^2)^2 - 4g^2 {F_2 F_2^*}}} \cr
& & - \frac{1}{2g|F_2|} (\Sigma^2 + 4g^2 |{z_2}|^2 + 2g \sqrt{F_2 F_2^*})
\ln {\frac{\Sigma^2 + 4g^2 |{z_2}|^2 + 2g \sqrt{F_2 F_2^*}}{\Sigma^2
+ 4g^2 |{z_2}|^2 - 2g \sqrt{F_2 F_2^*}}} \cr
& & - 8g^2 |z_2|^2 \ln {\frac{\Lambda^2}{4g^2 |z_2|^2 }}],
\end{eqnarray}
where $G = \frac{g^2}{m^2}$ and
\begin{eqnarray}
\frac{F_2}{Bz_2} + 1 = - \frac{NG}{16\pi^2} \frac{\Delta^2}{B} \frac{F_2}{z_2}
[ \ln{ \frac{\Lambda^4}{(\Sigma^2 + 4g^2 |{z_2}|^2)^2 - 4g^2 {F_2 F_2^*}}} \cr
- \frac{1}{2g|F_2|} (\Sigma^2 + 4g^2 |{z_2}|^2) \ln {\frac{\Sigma^2
+ 4g^2 |{z_2}|^2 + 2g \sqrt{F_2 F_2^*}}{\Sigma^2 + 4g^2 |{z_2}|^2
- 2g \sqrt{F_2 F_2^*}}}].
\end{eqnarray}

\vfill
\end{document}